# Development of Interatomic Potential for Al-Tb Alloy by Deep Neural Network Learning Method

L. Tang[1], Z. J. Yang[1,*], T. Q. Wen[2], K. M. Ho[2,3], M. J. Kramer[2] and C. Z. Wang[2,3,†]

[1]*Department of Applied Physics, College of Science, Zhejiang University of Technology, Hangzhou, 310023, China*
[2]*Ames Laboratory-USDOE, Iowa State University, Ames, Iowa 50011, USA*
[3]*Department of Physics and Astronomy, Iowa State University, Ames, Iowa 50011, USA*

Corresponding author: * zejinyang@zjut.edu.cn † wangcz@ameslab.gov

## Abstract

An interatomic potential for Al-Tb alloy around the composition of $Al_{90}Tb_{10}$ was developed using the deep neural network (DNN) learning method. The atomic configurations and the corresponding total potential energies and forces on each atom obtained from *ab initio* molecular dynamics (AIMD) simulations are collected to train a DNN model to construct the interatomic potential for Al-Tb alloy. We show the obtained DNN model can well reproduce the energies and forces calculated by AIMD. Molecular dynamics (MD) simulations using the DNN interatomic potential also accurately describe the structural properties of $Al_{90}Tb_{10}$ liquid, such as the partial pair correlation functions (PPCFs) and the bond angle distributions, in comparison with the results from AIMD. Furthermore, the developed DNN interatomic potential predicts the formation energies of crystalline phases of Al-Tb system with the accuracy comparable to *ab initio* calculations. The structure factor of $Al_{90}Tb_{10}$ metallic glass obtained by MD simulation using the developed DNN interatomic potential is also in good agreement with the experimental X-ray diffraction data.

## I. Introduction

Aluminum-rare-earth (Al-RE) binary alloys with Al-rich composition (about 90 at. % Al) can form metallic glasses by rapid quenching from the liquid state [1]. It has been shown that these Al-RE alloys have very high strength-to-weight ratio owing to a high proportion of light weight Al [2-4]. However, Al-RE alloys belong to marginal glass-forming systems which usually have high density of nanocrystals in the samples prepared by rapid solidification process [1, 5]. Moreover, the stable as well as metastable Al-rich intermetallics vary across lanthanide series [1, 6-8].

In order to understand the microscopic mechanisms of phase formation and competition in these complex alloys, the knowledge of short to medium range structural orders in liquid and undercooled liquid at atomistic level and the corresponding time evolution of the atomistic structures during solidification/devitrification are critical. Investigation of the atomistic structural difference among these binary alloys at liquid, glass as well as crystalline phases will provide us valuable insights to further tune these alloys for better properties and glass-forming ability.

We note that while most interest in the literature has been focused on light RE (e.g., Al-Ce [8]), Al alloyed with heavy RE have not been extensively investigated, except for Al-Sm system at the composition around Al-90 at. % where both experimental studies and MD simulations using an empirical interatomic potential have been reported [9-11]. One of the bottlenecks hindering computational simulation of Al-RE alloys is that reliable interatomic potentials for the most of Al-RE alloys (e.g., Al-Tb) are still lacking. Although *ab initio* calculations can offer high accuracy of interatomic interactions for Al-RE alloys, it can deal with only up to several hundred atoms and within a few hundred picoseconds (ps) simulation time in most of the simulations due to the expensive computational cost of the method. Therefore, it is difficult for AIMD simulations to investigate the long-time relaxation in glass and phase competition during solidification, which is a key to understand the metallic glass formation.

Recently, a deep learning method [12-14] with DNN model for many-body interatomic interactions has been developed which is very promising for overcoming the dilemma in simulation speed and accuracy. In a recently developed DNN learning software package called DeePMD-kit [14], the snapshots (which include the total potential energies, forces on each atom, and virial for a set of atomic configurations) from *ab initio* calculations are used to train interatomic potentials through DNN machine learning. After the training process, the obtained DNN model is not only able to accurately reproduce the potential energies and forces in the training data set, but also accurately predicts structural and dynamical properties of the materials being modeled. These advantages make the DNN learning method suitable for studying solidification and devitrification of alloy systems even the phase competition and transition in these systems are complex. Moreover, the interatomic potential constructed by the trained DNN model can be ready to use in standard LAMMPS package [15] to perform MD simulations. The computational cost of the MD simulations with DNN interatomic potentials scales linearly with system size, which enables us to investigate the long-range correlations and long-time relaxations in metallic glass systems.

In this paper, development of interatomic potential for Al-rich Al-Tb Alloys by the DNN learning method is presented. In order to enhance the sampling space, the training data set for the DNN model include snapshots from pure Al and Tb liquids as well $Al_{90}Tb_{10}$ liquid at various temperature, in addition to various crystalline phases of pure Al, pure Tb, and Al-Tb binary compounds (see Table. 1). The potential energies and forces in the training data set are calculated by first-principles density functional theory (DFT) using VASP [16, 17]. We demonstrate that the obtained DNN interatomic potential from the machine learning describes accurately the structures of $Al_{90}Tb_{10}$ liquid/glass and various Al-Tb crystalline phases in comparison with those from *ab initio* calculations and experiments.

The rest of paper is organized as follows. The DNN learning method for interatomic potential will be described in Sec. II. In Sec. III, we will present the details of data set generation for DNN model training and the parameters of DNN learning process. In Sec. IV, we will demonstrate the reliability of obtained DNN interatomic potential. Finally, summaries and conclusions are given in Sec. V.

## II. Interatomic potential by deep neural network learning method

Artificial neural networks (NNs), inspired by the biological NNs that constitute human brain, provide an accurate tool for the representation of arbitrary functions. A NN contains interconnected layers of nodes. There are three essential types of layers: an input layer, an output layer and hidden layers (which can be multilayers depending on the complexity of the model). The input layer collects input patterns. Hidden layers perform the learning functions by adjusting the network parameters to minimize the lost function defined in the NN model. The output layer has classifications or output signals to which input patterns may map. A "node" in a NN is a mathematical function that collects and classifies information according to a specific architecture. To model the interatomic potential by NN, the information fed to the input layer is a set of descriptors $\{D_i\}$ which describe atomistic environment around every atom $i$ of the structures in the training data set. The information extracted from the output layer are the energy $E_i$ on each atom. Then the total potential energy $E$ of each structure can be written as the sum of atomic energy $E_i$, i.e., $E = \sum_i E_i$. The mapping from the local environment of atom $i$ (i.e., $\{D_i\}$) to energy of each atom $E_i$ is done by the hidden layers in the NN model where the connection weights between the nodes in different layers and the bias parameter on each node of the hidden layers are used to model this mapping [13, 18]. These weights and bias parameters are obtained by NN training which optimize the lost function with respect to the training data set. Therefore, the potential energy surface of the system is acquired once the parameters in NN have been determined by training process. Furthermore, forces on each atom can be calculated analytically from the potential energy represented by NN. A schematic illustration of artificial NNs for modeling interatomic potentials is shown in Fig. 1.

In the present study, the Potential-Smooth version of DeePMD-kit software package [13] is used to train the DNN interatomic potentials. It has demonstrated this deep learning method is very robust in developing interatomic potentials for MD simulation studies of liquid, crystalline bulk structures and organic molecules [13, 19].

A crucial step in modeling interatomic potential by DNN is the construction of local structure descriptor $\{D_i\}$ from the Cartesian coordinates of the input atomistic structures. To ensure the invariance of the total energy with respect to rotation or translation of the structures or the interchanging of two atoms of the same element in the structure, the descriptors $\{D_i\}$ have to satisfy such invariance conditions. In early work of Behler and Parrinello, a set of symmetry functions has been used for the descriptors [18]. In the present work, we adopt the local coordinate frame developed in DeePMD-kit [14] to construct the $\{D_i\}$. The description of the local environment of atom $i$ is constructed in two steps. First, the relative Cartesian coordination $\{R_i\}$ of neighboring atoms $j$ within cutoff radius $r_c$ with respect to atom $i$ are transferred to the generalized coordination $\{\tilde{R}_i\}$ as

$$\{R_i\} = \{x_{ji}, y_{ji}, z_{ji}\} \rightarrow \{\tilde{R}_i\} = \{s(r_{ji}), \hat{x}_{ji}, \hat{y}_{ji}, \hat{z}_{ji}\} \quad (1)$$

where $\hat{x}_{ji} = s(r_{ji})x_{ji}/r_{ji}$, $\hat{y}_{ji} = s(r_{ji})y_{ji}/r_{ji}$, and $\hat{z}_{ji} = s(r_{ji})z_{ji}/r_{ji}$ have angular information of local environment. The radial information is in $s(r_{ji})$ which is smooth at the boundary of cutoff radius $r_c$. It is defined as [13]

$$s(r_{ji}) = \begin{cases} \frac{1}{r_{ji}}, \; r_{ji} < r_{cs} \\ \frac{1}{r_{ji}}\left\{\frac{1}{2}\cos\left[\pi\frac{r_{ji}-r_{cs}}{r_c-r_{cs}}\right] + \frac{1}{2}\right\}, r_{cs} < r_{ji} < r_c \\ 0, \; r_{ji} > r_c \end{cases} \quad (2)$$

where $r_{cs}$ is smooth cutoff parameter. Second, an embedding neural network (called filter NN) is introduced, where the radial information $s(r_{ji})$ are fed to its input. The output of filter NN will serve as weight coefficients to the generalized coordination $\{\tilde{R}_i\}$ in constructing the local structure descriptor $\{D_i\}$ which describes the local environment of atom $i$. Finally, the local structure descriptor $\{D_i\}$ is fed to the input of another neural network (called fitting NN), yielding the atomic energy $E_i$, thus the mapping from local configuration to atomic energy is achieved.

The training process is a procedure of optimizing the parameters in filter and fitting NNs by deep learning software package such as TensorFlow [20] to minimize the total loss function. In the present work, the total loss function $L = \frac{1}{S_b}\sum_{k=1}^{S_b} L_k$ is evaluated on each training step for a subset of training data (called a batch), where $S_b$ is the total number of snapshots in the batch. $L_k$ is the loss function for the *k*th snapshot in the batch and is defined as

$$L_k = \frac{p_e}{N}\left|\Delta E^{(k)}\right|^2 + \frac{p_f}{3N}\sum_i\left|\Delta f_i^{(k)}\right|^2 \qquad (3)$$

where the total potential energy error $\Delta E^{(k)}$ and force error $\Delta f_i^{(k)}$ on atom $i$ are the differences between the DNN predictions and *ab initio* calculation results for the atomic structure of the *k*th snapshot, respectively. *N* is total number of atoms in the structure. $p_e$ and $p_f$ are prefactors for energy and force respectively, which are continuously changing during the training process for optimization of DNN.

## III. Training data preparation and training process

The training data set is critical to the success of the NN machine learning to generate accurate interatomic potentials for reliable MD simulations. The target of our DNN model is to simulate the liquid and glass structures of $Al_{90}Tb_{10}$ alloy. Hence, the training data set is primarily composed of the snapshots of liquid $Al_{90}Tb_{10}$ at different temperatures prepared by AIMD simulations. The AIMD simulations were performed using a cubic cell containing 180 Al atoms and 20 Tb atoms and with periodic boundary conditions. The size of unit cell is $15.989\text{Å} \times 15.989\text{Å} \times 15.989\text{Å}$, which is chosen according to the density of liquid Al ($n_{\text{Al}}$) and liquid Tb ($n_{\text{Tb}}$), i.e., $n_{\text{Al}_{90}\text{Tb}_{10}} = 0.9n_{\text{Al}} + 0.1n_{\text{Tb}}$.

All the energies and forces of the structures in the training and validation data sets described below are calculated by VASP package. The time step of AIMD is taken as 3fs and NVT ensemble with Nose-Hoover thermostat [21, 22] are used in all simulations. The projector-augmented-wave (PAW) method [23] is used to describe the core-valence electron interaction. The generalized gradient approximation (GGA) in the Predew–Burke–Ernzerhof (PBE) form [24] is used for the electronic exchange and correlation potential. The default energy cutoff for the plane wave basis set from PBE potential is used and only the gamma point is used to sample the Brillouin zone in all AIMD simulations.

The initial configuration for the AIMD simulations was randomly selected from those generated by classical MD simulations using the available interatomic potential for Al-Sm [9]. Then the Sm atoms are replaced by Tb atoms and AIMD simulation was performed at 2000K for 2000 MD steps. Next, the sample is cooled down to 800K at a cooling rate of $3.3 \times 10^{13}$ K/s. During this cooling process, snapshot atomic configurations at the temperatures of 2000K, 1800K, 1600K, 1400K, 1200K, 1100K, 1000K, 900K, 800K, respectively, are randomly picked up to initialize the isothermal MD simulations at the corresponding temperatures. The isothermal MD simulations for each temperature was performed for 90 ps and snapshots at every step of the AIMD simulations are collected. The total number of the snapshots collected for the 9 temperatures are 270,000, among them 240,000 are randomly selected as training data set and the rest of 30,000 are used as validation data set for testing the trained DNN model.

In experiment, phase separation of fcc Al has been observed in the as-quenched Al-Tb glass [25]. In order to ensure that the DNN potential can handle correctly possible phase separation, we also add snapshots of the pure Al and Tb liquids/solids into the training data set. Both the pure liquid Al (Tb) and crystal fcc Al (hcp Tb) are calculated by VASP and included in the training data set. For the pure liquid Al, the sample is simulated isothermally at T=1400K and 2000K while the simulation temperatures for liquid Tb is 1800K and 2200K, respectively. Both Al and Tb liquid sample contain 108 atoms and with periodic boundary conditions. The AIMD simulations of Al (Tb) liquid are first performed isothermally at liquid phase temperatures for 2000 steps to melt the samples and obtain the liquid state of Al (Tb). After that, all the MD steps during the following 30ps simulations for each temperature are collected. Then, in all the *ab initio* data 18,000 snapshots for Al or Tb liquids respectively are randomly picked up for the training data set. In addition, the remaining 2,000 snapshots of each pure liquids are collected for the validation data set.

The training and validation data set for our DNN learning model also included the information of the relevant crystalline phases. In order to obtain the snapshots of crystalline phases, AIMD simulations with a supercell of fcc Al (hcp Tb) containing 108 atoms at finite temperature are employed. It allows the atoms to move around the equilibrated positions in the crystals and then generate a serial snapshots of crystal structures with distortions. Moreover, in order to obtain the information about atomic structures and forces far from the equilibrium, we also carry out the *ab initio* simulations at different lattice constants. For fcc Al crystal the lattice constant is $a = 4.05(1 \pm 0.02n)$ Å, $n = 0,1,2,3,4,5$. For hcp Tb crystal the lattice constant is $a = 3.60(1 \pm 0.02n)$ Å, $c = 5.70(1 \pm 0.02n)$Å, $n = 0,1,2,3,4,5$. At each lattice constant, all atoms are distorted by means of AIMD simulations at T=300K. In these ways, we generate and randomly select 2,000 distorted fcc Al crystal structures and 2,000 distorted hcp Tb crystal structures at different lattice constants (or pressures) to be included in the training data set. Another 200 such distorted Al and Tb crystal structures respectively are also collected in the validation data set. In addition to the pure Al and Tb crystalline phases, we also include the known crystalline phases of Al-Tb alloy which covered the whole composition range, i.e., $Al_{17}Tb_2$ [26], $Al_4Tb$ [27], $Al_3Tb$ [28], $Al_2Tb$ [29], AlTb [30], $Al_2Tb_3$ [31], $Al_2Tb$ [32] and $AlTb_3$ [33], to the training data set. For each of these crystalline phases, the snapshot structures were generated in the same way as that used for Al and Tb crystalline structures described above. Similarly, 2000 snapshot structures from AIMD simulations for each compound are included in the training data set and another 200 snapshots are used for the validation data set. The overall information of training and validation data set are summarized in Table 1.

In the DNN training process, in order to capture the local configuration information up to the second neighboring shell, the radial cutoff $r_c$ is taken as 7.2 Å which is about the radial of second shell from the PPCFs in AIMD simulations of $Al_{90}Tb_{10}$ liquid. The smooth cutoff parameter $r_{cs}$ is chosen to 7.0 Å. The filter neural network has two hidden layers with 50 and 100 nodes, respectively. The fitting neural network model has three hidden layers with equal numbers of nodes (240 nodes) per layer. The DNN is initialized with random numbers and the total number of training steps is 2,000,000. The exponentially decaying learning rate is used. At the *i*th training step the learning rate is defined as $r_l(i) = 0.001 \times 0.96^{i/10000}$, where the start learning rate is 0.001 and the decay rate is 0.96 with decay step of 10000. The energy prefactor $p_e$ in loss function starts at 0.2 and ends up to 2. Meanwhile, for forces the prefactor $p_f$ is 100 at beginning and goes down to 1 at the end of training process.

## IV. Performance of the deep neural network interatomic potential

Fig. 2(a)-(d) show the comparison of energies and forces from the trained DNN model and *ab initio* results for 1,000 snapshots of $Al_{90}Tb_{10}$ liquid which are randomly picked up from the training and validation data set, respectively. The vertical coordinate represents the energies (or forces along *x* axis) of the snapshot structures calculated by the trained DNN model while the horizontal coordinate is the corresponding energies (or forces along *x* axis) obtained by *ab initio* calculations. It can be seen that the trained DNN model not only well reproduces the *ab initio* results in the training data set but also accurately predict the energies and forces for the snapshots in the validation data set. The root mean square (RMS) error of energy is below 3.0 meV/atom and the force RMS error is on the order of 0.1 eV/Å, which is sufficient for investigating the structures and dynamics of liquid. Moreover, the trained DNN model can also well predict the energies and forces for the atomic configurations which are not included in the training or validation data set. For example, although the snapshots in the AIMD simulation at temperature 1300K are not included in the training or validation data set, Fig. 2(e) and (f) show excellent prediction of energies and forces of these atomic configurations. More details of the energy and force RMS errors from the DNN model predictions for all the systems which are used to train DNN are shown Table. 1. It can be seen that the obtained DNN can well reproduce all the *ab initio* results including both liquid and crystalline structures.

The reliability and transferability of the obtained DNN potential are further tested by using it in a MD package such as LAMMPS to study the temperature dependent structures of liquid. Fig. 3 shows the comparison of the total PPCF, Al-Al, Al-Tb, and Tb-Tb partial PPCFs of liquid $Al_{90}Tb_{10}$ at T=1300K and 2000K calculated by AIMD and MD simulation with DNN potential. The initial configurations of both AIMD and MD with DNN potential is the same. The simulation times for statistical average of PPCF are 270ps and 30ps for the samples at T=1300K and 2000K, respectively. It shows that the PPCFs from MD with DNN potential agree well with those from AIMD simulations even the snapshots of the liquid at 1300K have not been included in the training data set. Note that some small differences in the Tb-Tb PPCF between AIMD and MD with DNN potential can be attributed to the relatively poor statistics due to the small number of Tb atoms (only 20 atoms in a box) used in the simulation. In addition, the MD with DNN potential can also well reproduce the PPCF of pure Al and Tb liquid in comparison with those from AIMD, as shown in the Fig. 4.

Besides the PPCF, the bond angle distribution can provide more structure information about the liquid samples. Thus, it can also be used to test the reliability of DNN potential. We take the first minima of PPCF as the cutoff distances to calculate the bond angle distributions and all the structures used for PPCF calculations in Fig. 3 are used to perform the statistics of bonding angles. The bond angle distributions for $Al_{90}Tb_{10}$ liquid at T=1300K and 2000K obtained in this way are shown in Fig. 5 and 6, respectively. It can be seen that the DNN potential can well reproduce the bond angle distributions from AIMD simulations.

Since glass formability strongly dependent on the competition with the nucleation and growth of various nearby crystalline phases, it is critical that the developed DNN potential can describe well the energy landscape of the Al-Tb system at the composition of interest including possible stable and metastable crystalline phases. The competition among these crystalline phases and the glass formation upon the solidification would highly correlates with their formation energies. Here the trained DNN potential is used to calculate the formation energies of crystalline phases in Al-Tb system at T=0K. The energies of pure fcc Al and pure hcp Tb are used as the reference for calculating the formation energy. The formation energy for crystalline phase $\mathrm{Al}_m\mathrm{Tb}_n$ is defined as $E_{\mathrm{form}}(\mathrm{Al}_m\mathrm{Tb}_n) = [E(\mathrm{Al}_m\mathrm{Tb}_n) - nE(\mathrm{Al}) - mE(\mathrm{Tb})]/(n+m)$ . Fig. 7 shows the comparison of formation energies between DNN potential and *ab initio* calculations. In both *ab initio* and DNN potential calculation, the conjugate gradient algorithm is used to optimize the atomic structures. It can be seen that the formation energies of known stable crystalline phases in training data set predicted by DNN potential agree well with the results of *ab initio* calculations. Besides the crystalline phases in the training data set, the obtained DNN potential can also well predict the formation energies of crystals that are not used for training DNN potential. For example, we calculated the formation energies of other two types of $Al_3Tb$ crystals (the hypothetical $Al_3Tb$ with structure of $Al_3Y$ [34, 35] and BaPb3-type $Al_3Tb$ founded in ref. 28). The results show that the DNN potential reproduced the order of formation energy for all the three $Al_3Tb$ phases. Recently, it is found that several complex metastable crystalline phases emerge in the devitrification process of $Al_{90}Sm_{10}$ system, and the structures of these complex phases have been identified by genetic algorithm (GA) search [36, 37]. These novel crystalline phases are valuable testing targets to validate the obtained DNN potential. We calculated the formation energies of $Al_{82}Tb_{10}$ (big tetra structure), $Al_{120}Tb_{22}$ (big cubic structure) and $Al_5Tb$ (big hex structure) metastable phases in which the Al composition is close to 90%. As shown in Fig. 7, the formation energies produced by the DNN potential are all close to the values of *ab initio* calculations. All the results of formation energies and relaxed lattice parameters obtained by DNN potential and *ab initio* calculations are listed in the Table. 2.

Finally, we also perform MD simulations of Al-Tb liquid and glass using the developed DNN potential with the number of atoms much larger than those structures in the training data set and compare the simulation results with the measurement from experiment. Experimentally, the liquid $Al_{91}Tb_9$ at 1174K was prepared by Cu-heart electric arc melting under Ar atmosphere and the glassy sample for $Al_{90}Tb_{10}$ was prepared by Cu block single melt-spinning technique, which were reported in ref. 25. The structure factors of the liquid and glass have been measured using high energy X-Ray diffraction (XRD) [25]. For comparison, our MD simulation with DNN potential of liquid $Al_{91}Tb_9$ is performed with 5000 atoms (4550 Al and 450 Tb) in a cubic box. The initial configuration is randomly picked up from the snapshots in previous MD simulation of liquid Al-Sm, where the Sm atoms are replaced by Tb atoms. First, the sample is equilibrated at 1174K for 30ps. Then MD trajectories in the subsequent simulation of 30ps at the same temperature are collected to calculate S($q$). Fig. 8(a) shows the calculated and experimental total structure factor of $Al_{91}Tb_9$ liquid at T=1174K. As one can see in Fig. 8(a), the first and second peaks of S(q) from MD with DNN potential agree well with the experimental data, except that there are some deviations around the first minimum. In addition, the height of pre-peak from MD with DNN potential is higher than that of experimental data. The glass MD sample of $Al_{90}Tb_{10}$ (4500 Al and 500 Tb) at T=300K is obtained by quenching from liquid of 2000K with cooling rate of $10^{11}$ K/s. One can see that the position and height of pre-peak from MD with DNN potential at T=300K agrees well with the experimental result, as shown in Fig. 8(b). Other peak positions and heights also agree well with experimental measurement. These results show that the developed DNN potential is suitable for MD simulations of Al-rich Al-Tb liquids and glasses.

## V. Summary

In this paper, we have developed a DNN interatomic potential for Al-rich Al-Tb alloys by DeePMD-kit software package based on deep learning method. The VASP package is used to calculate the snapshots of liquid and crystal Al-Tb structures to prepare the training data for machine learning. In order to train a transferable model, not only liquid $Al_{90}Tb_{10}$ but also the liquid of pure Al and pure Tb, as well as the crystalline structures of Al, Tb and binary Al-Tb compounds are included in the training data set to extend the sampling space. After the training process, the obtained DNN model has been demonstrated to predict accurately the energies and forces of Al-Tb system for both structures included and not included in the training data set.

The developed interatomic potential in the form of DNN model can be used in LAMMPS package to perform MD simulations. The results show that the DNN potential can well reproduce the PPCFs and bond angle distribution in AIMD simulations. Moreover, the calculated formation energies of crystalline phases of Al-Tb system using the DNN potential are found to be excellent agreement with *ab initio* results. Finally, the total structure factors of liquid and glass $Al_{90}Tb_{10}$ calculated by DNN potential agree well with the XRD data. In particular, the MD with DNN potential can well reproduce the positions and heights of the peaks in structure factors of $Al_{91}Tb_9$ liquid and $Al_{90}Tb_{10}$ glass as those measured in experiment. Our studies indicate that the developed DNN interatomic potential by deep learning method is reliable for MD simulation studies of Al-rich Al-Tb alloys, in both disordered liquid/glass and ordered crystalline compounds.

**Acknowledgements**

Work at Ames Laboratory was supported by the U.S. Department of Energy (DOE), Office of Science, Basic Energy Sciences, Materials Science and Engineering Division including a grant of computer time at the National Energy Research Supercomputing Center (NERSC) in Berkeley. Ames Laboratory is operated for the U.S. DOE by Iowa State University under contract # DE-AC02-07CH11358. L. Tang and Z. J. Yang acknowledge the support by the National Natural Science Foundation of China (Grant Nos. 11304279 and 11104247). Z. J. Yang also acknowledges the Natural Science Foundation of Zhejiang Province, China (Grant No. LY18E010007).

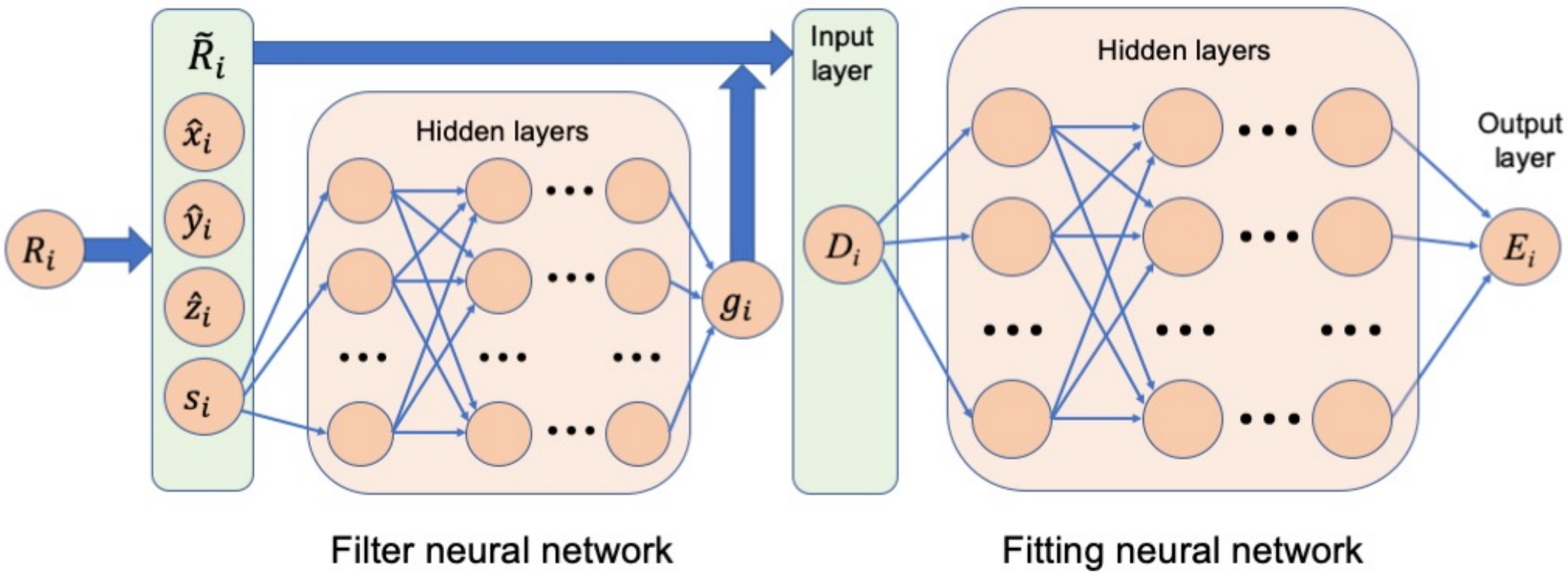

Fig. 1. The schematic illustration of deep learning method for modeling DNN interatomic potentials. First, the relative coordinates $\{R_i\}$ of neighboring atoms with respect to atoms $i$ within cutoff radius are converted to the generalized coordinates $\{\tilde{R}_i\}$, where $\{\hat{x}_i, \hat{y}_i, \hat{z}_i\}$ have angular information and $\{s_i\}$ has radial information of local atomic environment. Second, using the radial part $\{s_i\}$ in $\{\tilde{R}_i\}$ as input, the filter NN outputs the weight coefficients $\{g_i\}$ which are added to the generalized coordination $\{\tilde{R}_i\}$. Then, the local structure descriptor $\{D_i\}$ (preserves translation, rotation and permutation symmetries) which describes the local environment of atom $i$ is obtained. Next, $\{D_i\}$ enters into the fitting NN, yielding the atomic energy $E_i$ which is added to the total energy $E$.

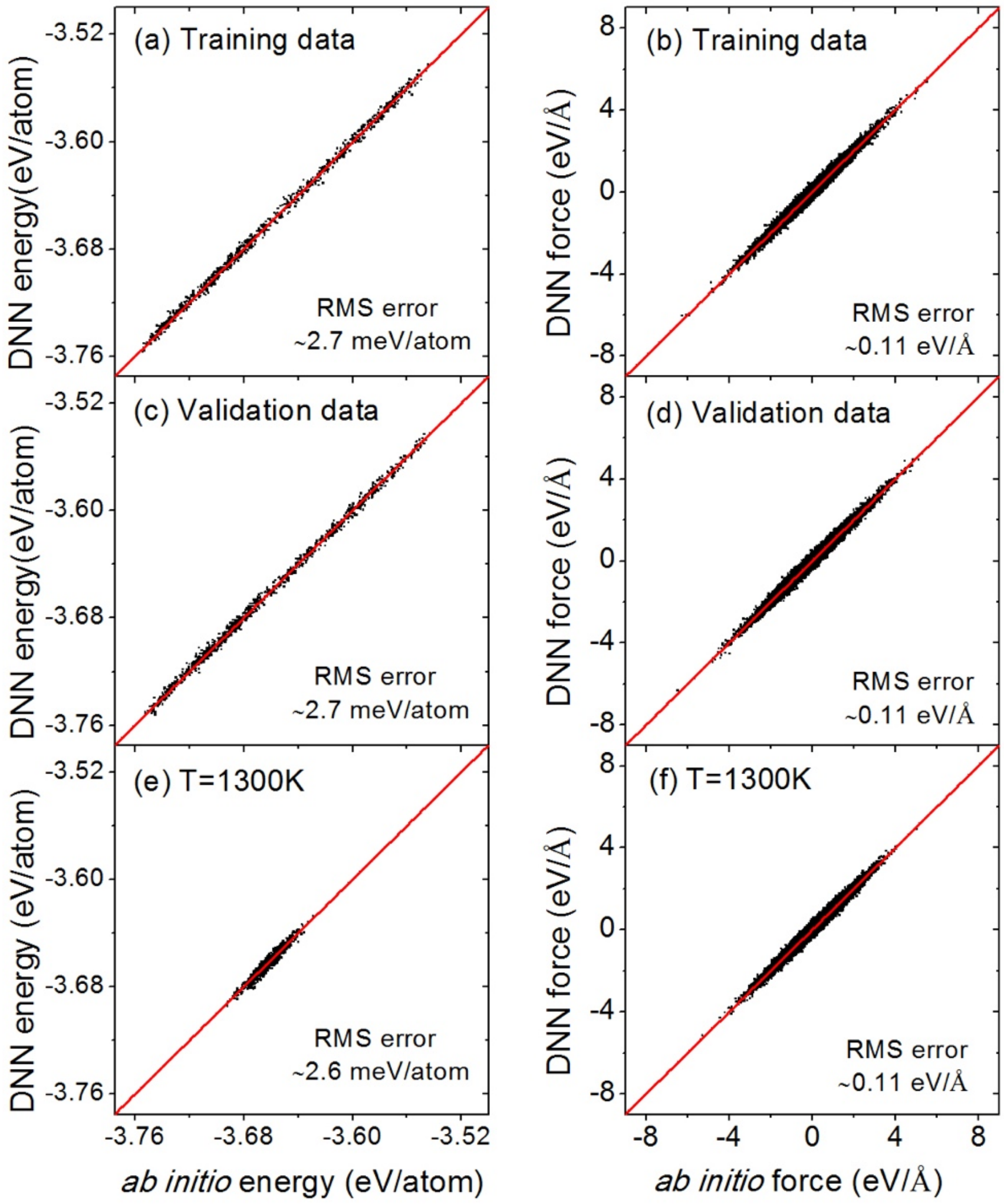

Fig. 2. Testing of energy and force predictions of the trained DNN model. Figure (a) and (b) are the comparisons of *ab initio* and DNN predicted energies and forces (along $x$ axis) on the 1000 snapshots of $Al_{90}Tb_{10}$ liquid which are randomly picked up from the training data set. In figure (c) and (d), the 1000 snapshots in the validation data set are randomly collected. To further test the performance of DNN predictions, in figure (e) and (f) the 1000 snapshots are collected from the AIMD simulation at T=1300K which is not included in the group of simulation temperatures for training or validation data set.

| Systems for training DNN | Total number of atoms in box/supercell | Simulation temperatures (K) | Total simulation time (ps) | Total number of snapshots in training data set | Total number of snapshots in validation data set | Energy error (meV/atom) | Force error (eV/Å) |
|---|---|---|---|---|---|---|---|
| $Al_{90}Tb_{10}$ liquid | 200 | 2000, 1800, 1600, 1400, 1200, 1100, 1000, 900, 800 | 90 for each temperature | 240,000 | 30,000 | 2.7 | 0.11 |
| Tb liquid | 108 | 1800, 2200 | 60 | 18,000 | 2,000 | 4.8 | 0.16 |
| Tb crystal | 108 | 300 | 6.6 | 2,000 | 200 | 4.4 | 0.09 |
| Al liquid | 108 | 1400, 2000 | 60 | 18,000 | 2,000 | 3.3 | 0.14 |
| Al crystal | 108 | 300 | 6.6 | 2,000 | 200 | 1.7 | 0.07 |
| $Al_{17}Tb_2$ | 304 | 300 | 6.6 | 2,000 | 200 | 3.0 | 0.08 |
| $Al_4Tb$ | 120 | 300 | 6.6 | 2,000 | 200 | 1.9 | 0.08 |
| $Al_3Tb$ | 240 | 300 | 6.6 | 2,000 | 200 | 1.4 | 0.08 |
| $Al_2Tb$ | 192 | 300 | 6.6 | 2,000 | 200 | 1.0 | 0.08 |
| AlTb | 64 | 300 | 6.6 | 2,000 | 200 | 3.0 | 0.10 |
| $Al_2Tb_3$ | 160 | 300 | 6.6 | 2,000 | 200 | 2.1 | 0.09 |
| $AlTb_2$ | 216 | 300 | 6.6 | 2,000 | 200 | 2.5 | 0.10 |
| $AlTb_3$ | 108 | 300 | 6.6 | 2,000 | 200 | 2.8 | 0.09 |

Table 1. The overall information of training and validation data set for Al-Tb system. The RMS errors of energy and force predicted by DNN model for the validation data set of various Al-Tb system are also shown in the table.

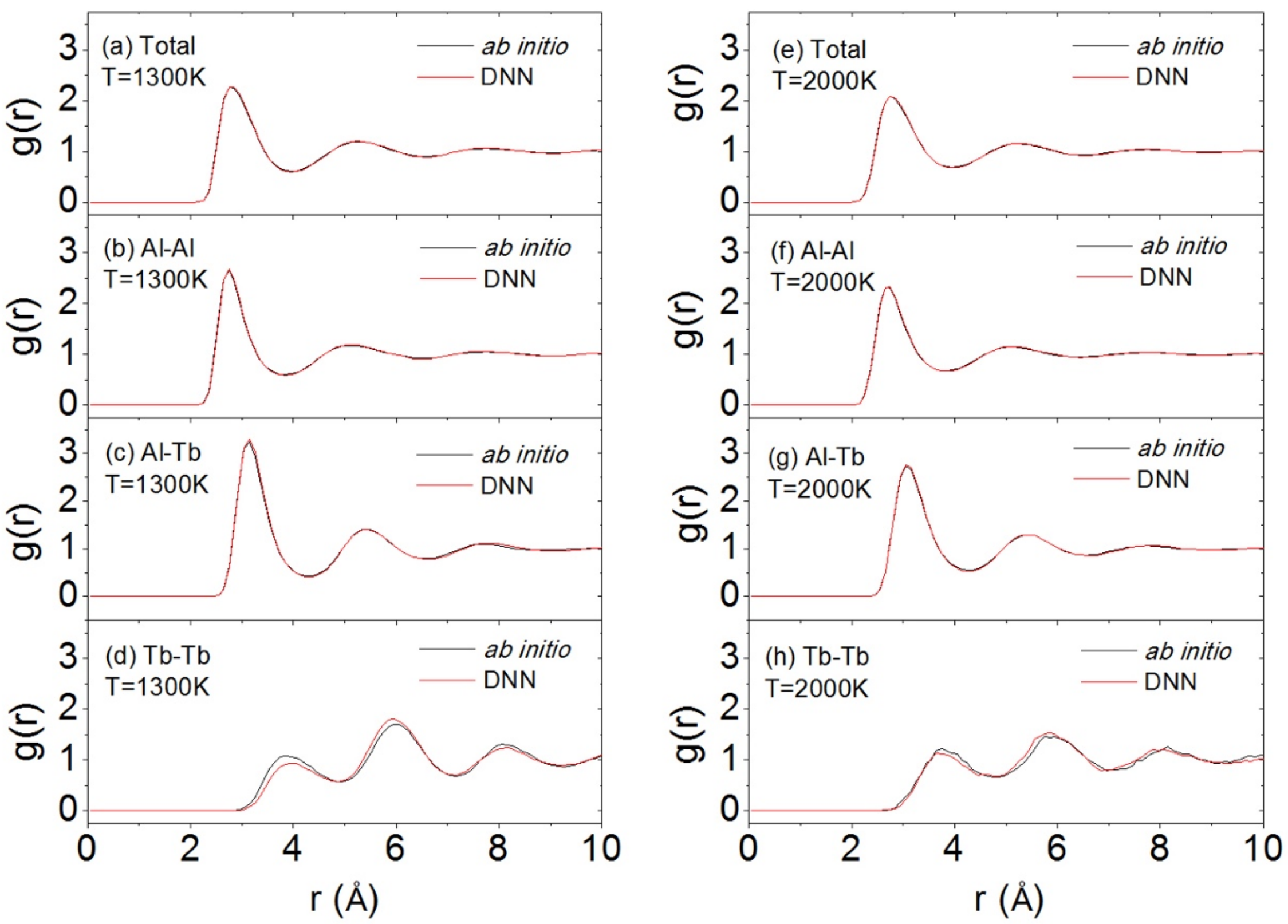

Fig. 3. Partial pair correlation functions in AIMD and MD simulations with DNN potential for liquid $Al_{90}Tb_{10}$ at (a)-(d)T=1300K and (e)-(h)2000K.

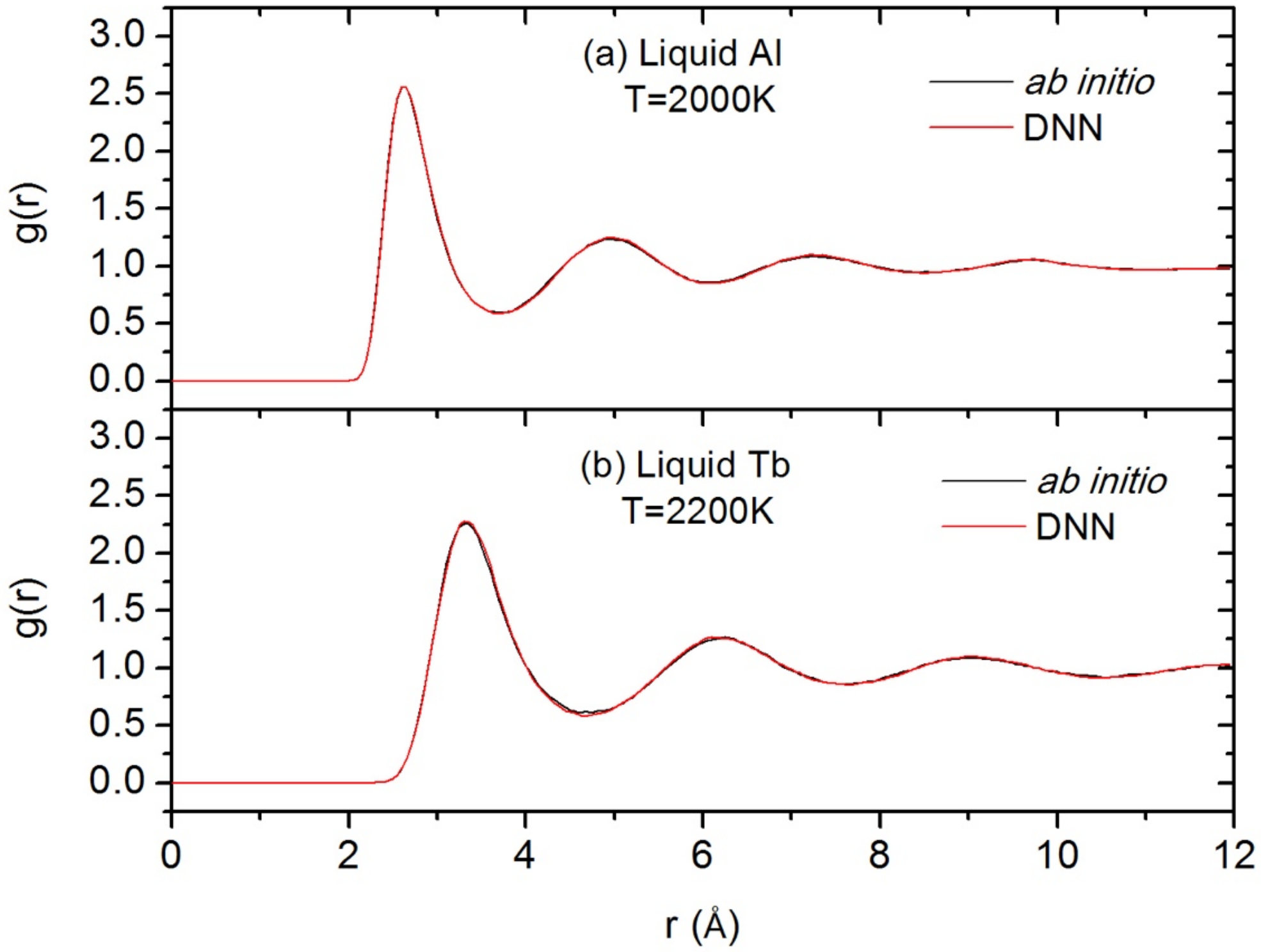

Fig. 4. Pair correlation functions in AIMD and MD simulations with DNN potential for pure liquid (a) Al and (b) Tb.

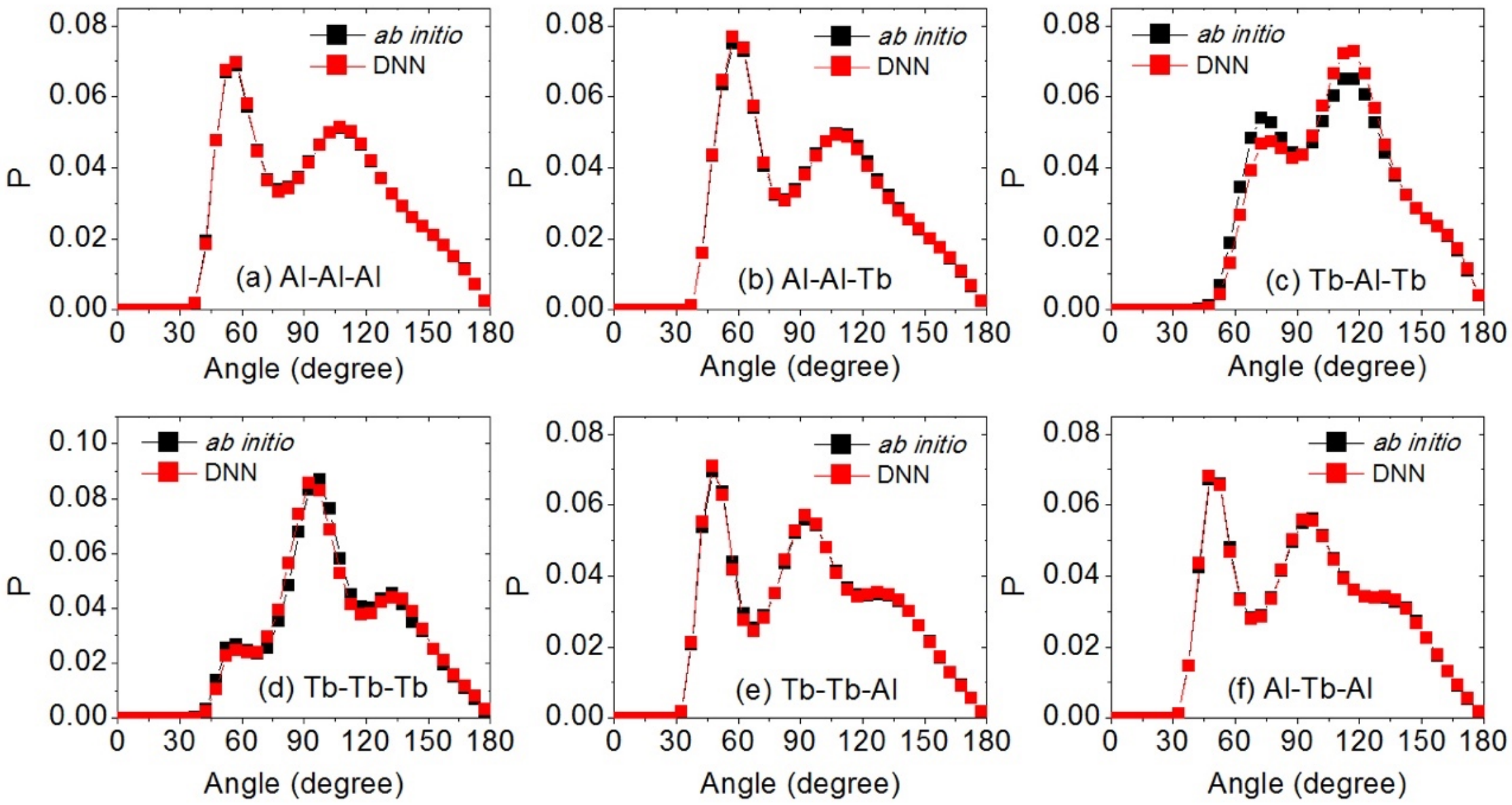

Fig. 5. The bond angle distributions in the liquid $Al_{90}Tb_{10}$ at T=1300K from AIMD and MD simulations with DNN potential.

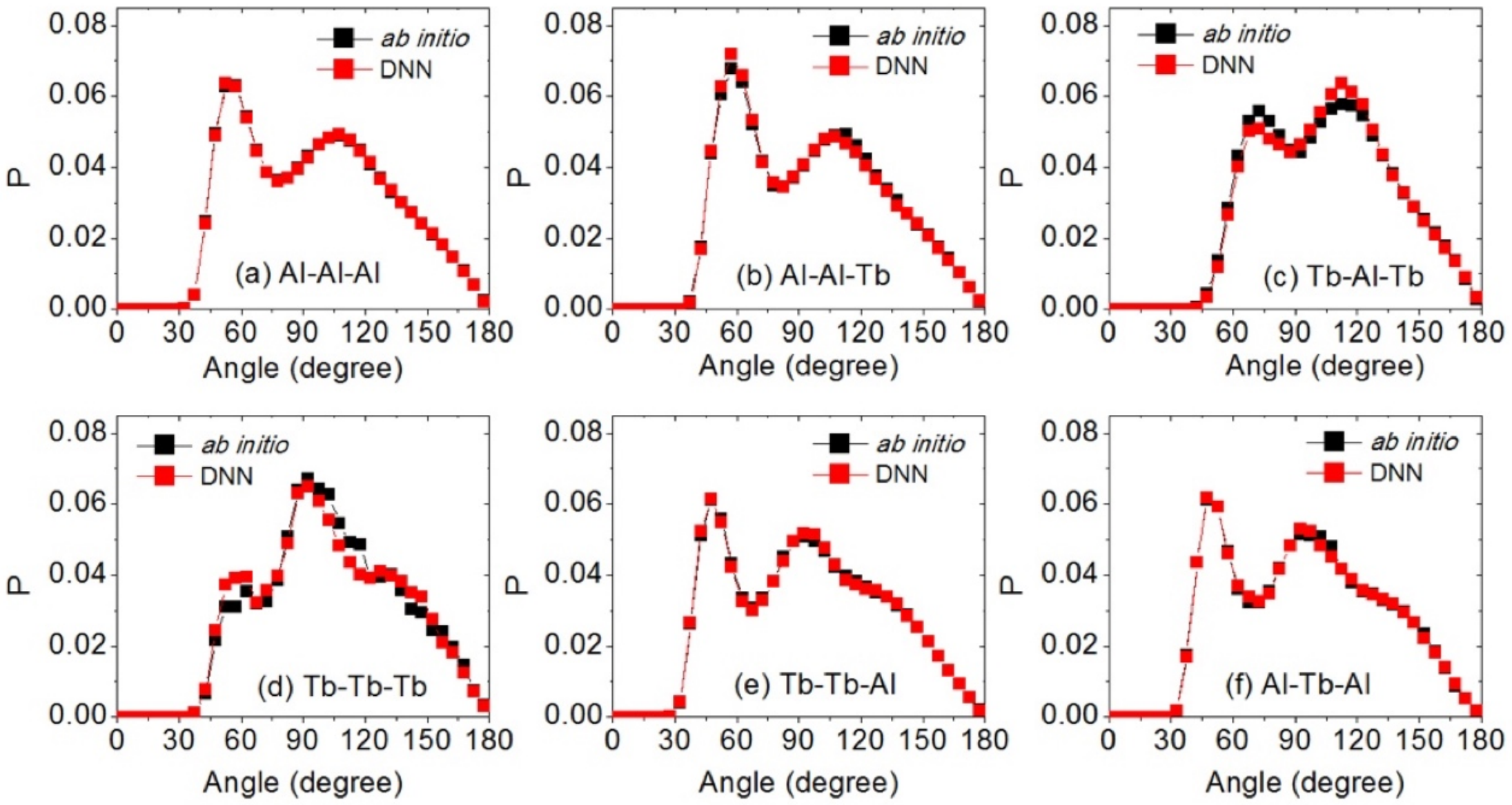

Fig. 6. The bond angle distributions in the liquid $Al_{90}Tb_{10}$ at T=2000K from AIMD and MD simulations with DNN potential.

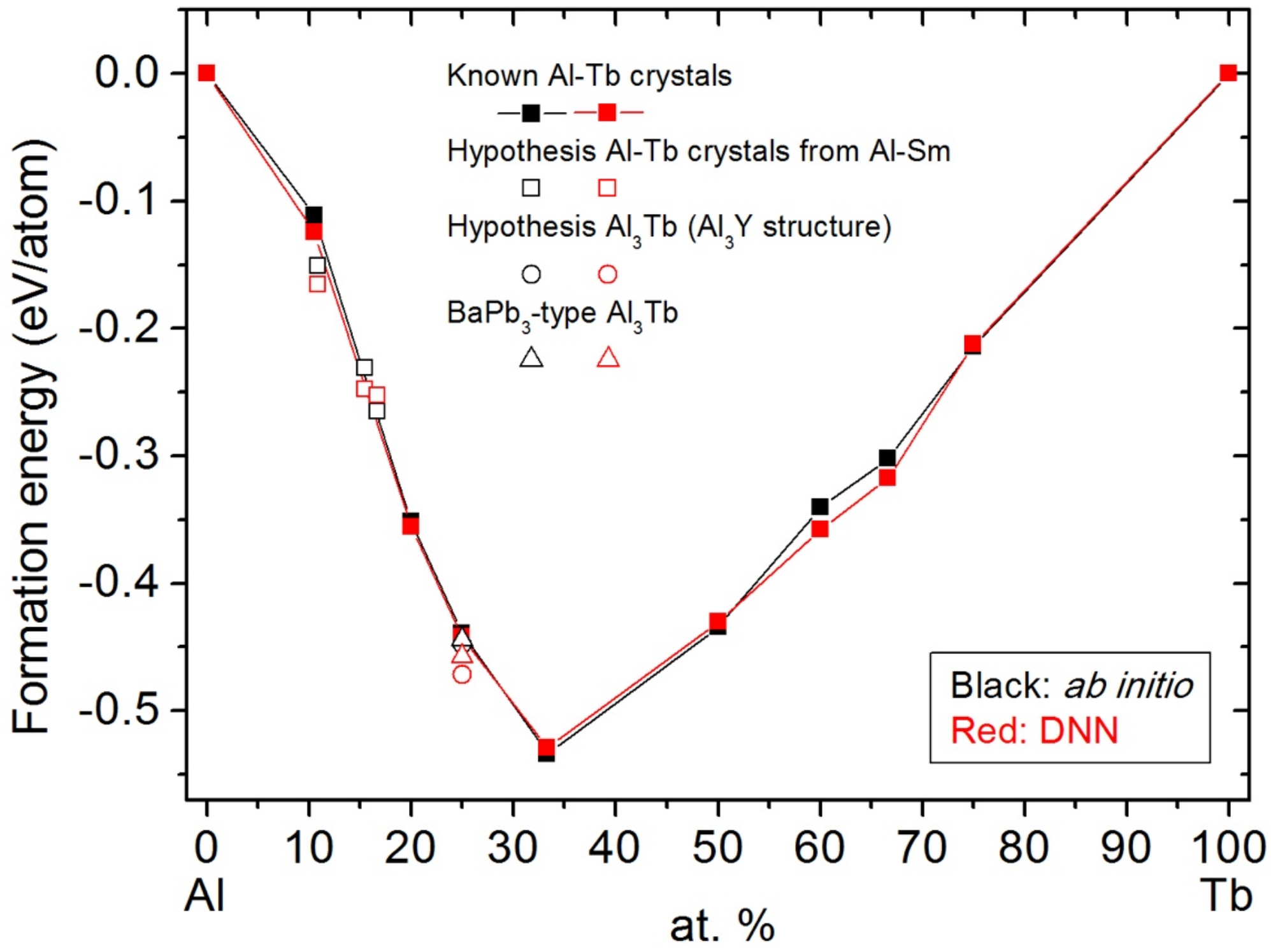

Fig. 7. The formation energies of Al-Tb system calculated by DNN potential and *ab initio* method at T=0K. The solid squares denote the known stable crystal structures in the training data set of Al-Tb system. The open squares denote the hypothesis Al-Tb crystal with the structures of metastable crystalline phases found in Al-Sm system. They are, from left to right, $Al_{82}Tb_{10}$ (big tetra structure), $Al_{120}Tb_{22}$ (big cubic structure), $Al_5Tb$ (big hex structure). The open circles and triangles denote the hypothetical $Al_3Tb$ with structure of $Al_3Y$ and the $BaPb_3$-type $Al_3Tb$ crystal, respectively. Noted that the open data is not included in the training data set, which suggests that the obtained DNN potential has ability to predict the formation energy for the unknown Al-Tb structure around composition of 10 at. % Tb.

| crystalline phase | Structure | a (Å) | b (Å) | c (Å) | $\alpha$ (°) | $\beta$ (°) | $\gamma$ (°) | Formation energy (eV/atom) |
|---|---|---|---|---|---|---|---|---|
| Al | 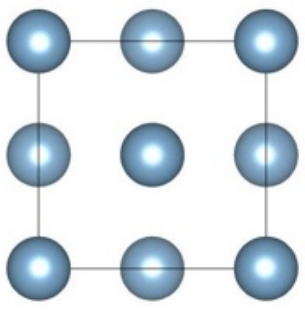 | 4.053 | 4.053 | 4.053 | 90 | 90 | 90 | 0 |
| | | 4.038 | 4.038 | 4.038 | 90 | 90 | 90 | 0 |
| $Al_{17}Tb_2$ | 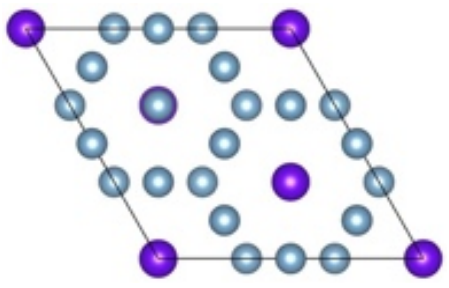 | 9.582 | 9.582 | 8.728 | 90 | 90 | 120 | -0.124 |
| | | 9.423 | 9.423 | 9.003 | 90 | 90 | 120 | -0.112 |
| $Al_4Tb$ | 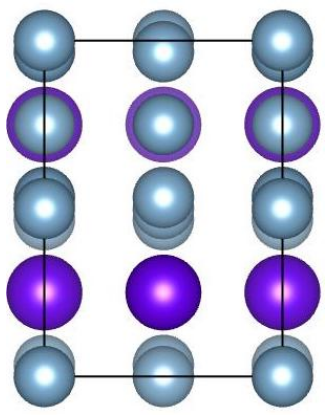 | 4.463 | 6.308 | 13.808 | 90 | 90 | 90 | -0.356 |
| | | 4.415 | 6.295 | 13.785 | 90 | 90 | 90 | -0.352 |
| $Al_3Tb$ | 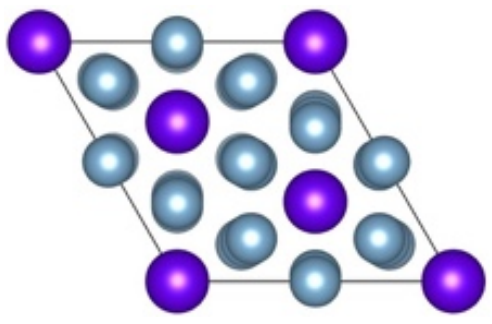 | 6.110 | 6.110 | 36.049 | 90 | 90 | 120 | -0.442 |
| | | 6.130 | 6.130 | 35.985 | 90 | 90 | 120 | -0.439 |
| $Al_2Tb$ | 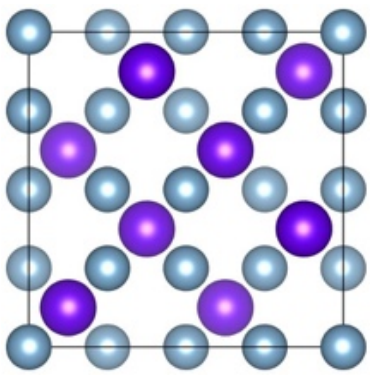 | 7.920 | 7.920 | 7.920 | 90 | 90 | 90 | -0.529 |
| | | 7.888 | 7.888 | 7.888 | 90 | 90 | 90 | -0.534 |
| AlTb | 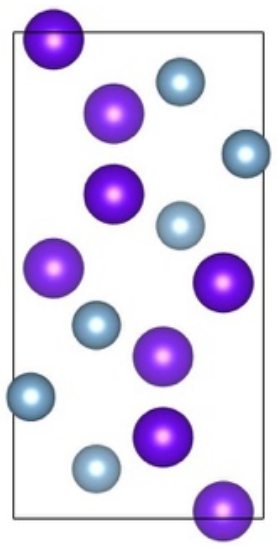 | 5.873 | 11.454 | 5.661 | 90 | 90 | 90 | -0.430 |
| | | 5.861 | 11.476 | 5.638 | 90 | 90 | 90 | -0.434 |
| $Al_2Tb_3$ | 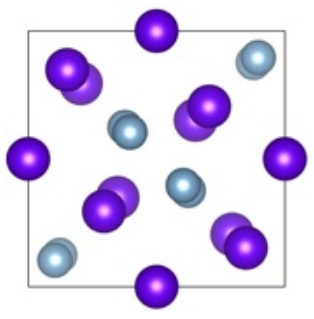 | 8.297 | 8.297 | 7.607 | 90 | 90 | 90 | -0.358 |
| | | 8.276 | 8.276 | 7.615 | 90 | 90 | 90 | -0.341 |
| $AlTb_2$ | 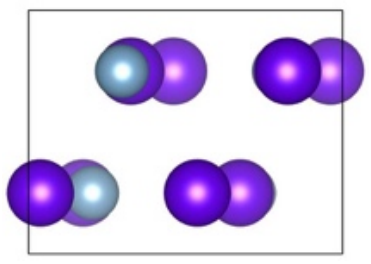 | 6.597 | 5.117 | 9.447 | 90 | 90 | 90 | -0.318 |
| | | 6.575 | 5.037 | 9.640 | 90 | 90 | 90 | -0.302 |
| $AlTb_3$ | 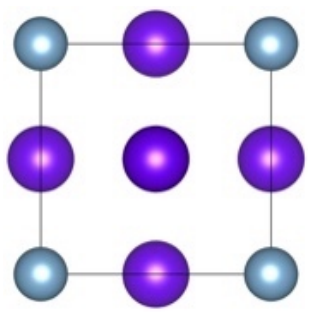 | 4.762 | 4.762 | 4.762 | 90 | 90 | 90 | -0.212 |
| | | 4.774 | 4.774 | 4.774 | 90 | 90 | 90 | -0.214 |
| Tb | 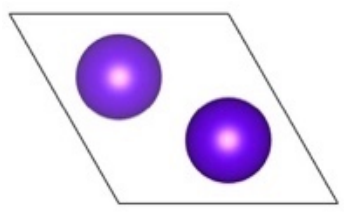 | 3.605 | 3.605 | 5.706 | 90 | 90 | 120 | 0 |
| | | 3.617 | 3.617 | 5.668 | 90 | 90 | 120 | 0 |
| $Al_{82}Tb_{10}$ (big tetra) | 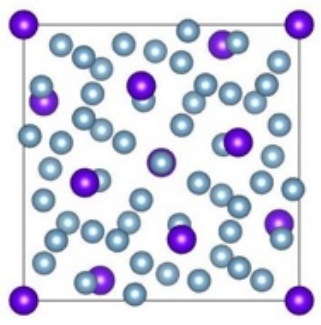 | 13.202 | 13.202 | 9.502 | 90 | 90 | 90 | -0.165 |
| | | 13.247 | 13.247 | 9.512 | 90 | 90 | 90 | -0.151 |
| $Al_{120}Tb_{22}$ (big cubic) | 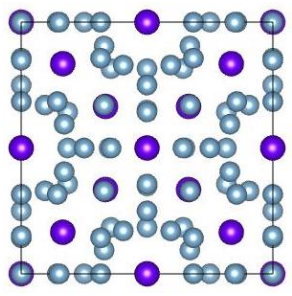 | 13.781 | 13.781 | 13.781 | 90 | 90 | 90 | -0.248 |
| | | 13.822 | 13.822 | 13.822 | 90 | 90 | 90 | -0.231 |
| $Al_5Tb$ (big hex) | 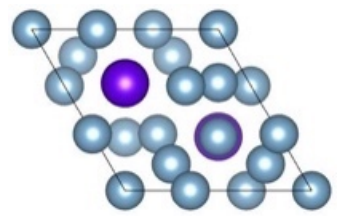 | 5.407 | 5.407 | 17.729 | 90 | 90 | 120 | -0.253 |
| | | 5.430 | 5.430 | 17.636 | 90 | 90 | 120 | -0.265 |
| $Al_3Tb$ ($Al_3Y$ structure) | 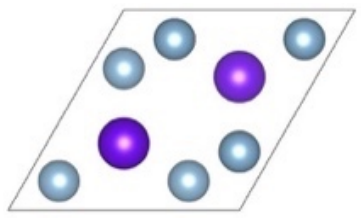 | 6.250 | 6.250 | 4.587 | 90 | 90 | 120 | -0.472 |
| | | 6.300 | 6.300 | 4.618 | 90 | 90 | 120 | -0.448 |
| $BaPb_3$-type $Al_3Tb$ (from ref. 28) | 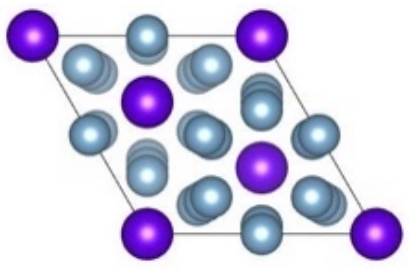 | 6.175 | 6.175 | 21.172 | 90 | 90 | 120 | -0.458 |
| | | 6.208 | 6.208 | 21.187 | 90 | 90 | 120 | -0.444 |

Table. 2. Lattice parameters and formation energies of Al-Tb crystalline phases. In calculation of the formation energy, fcc Al and hcp Tb crystal were taken as the reference states. The top value is reproduced by the DNN potential and the bottom one is calculated by the *ab initio* method.

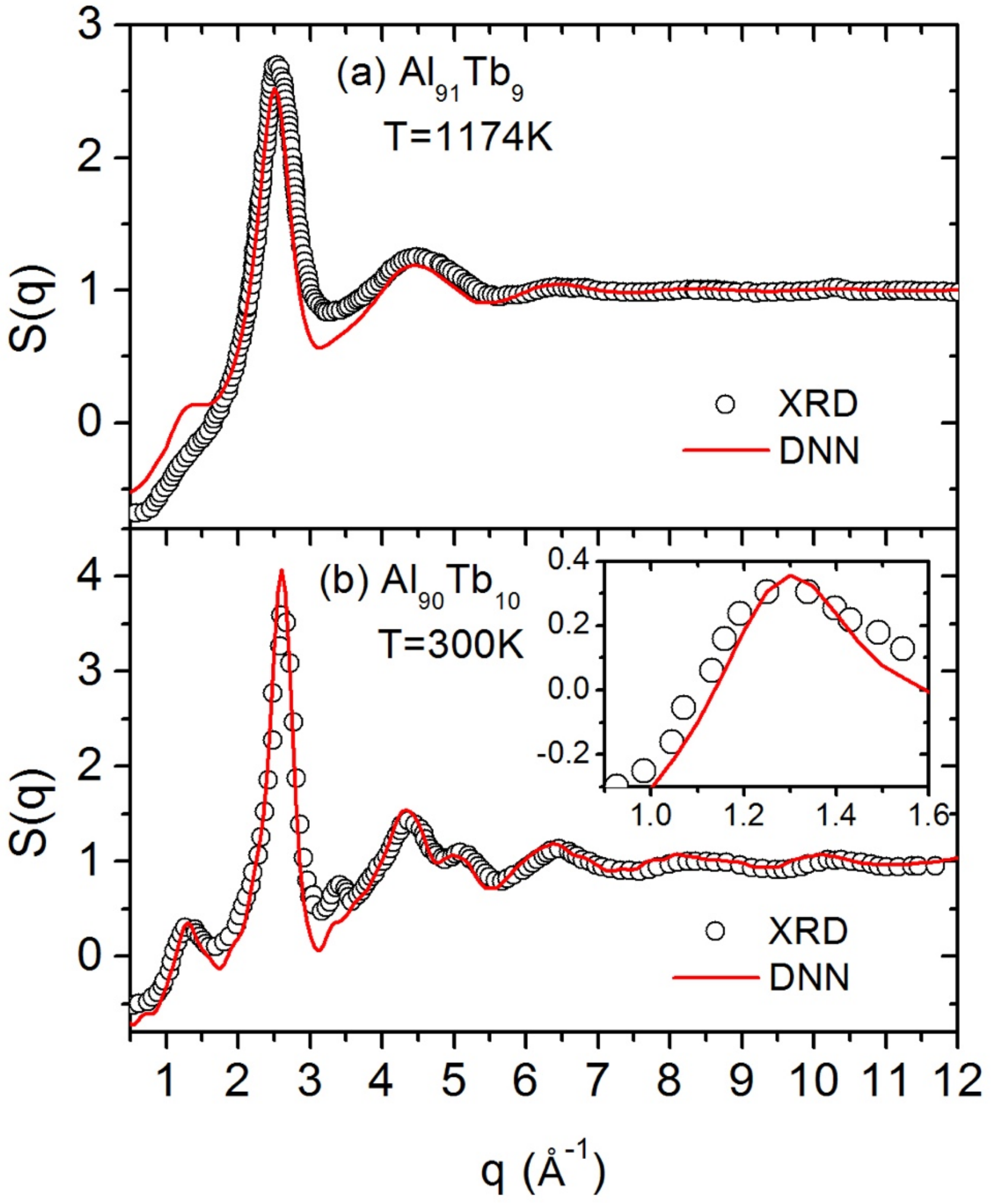

Fig. 8. The total structure factor of (a) liquid $Al_{91}Tb_9$ and (b) amorphous $Al_{90}Tb_{10}$ alloy. To obtain the amorphous $Al_{90}Tb_{10}$ for simulation, the sample is quenched from liquid at cooling rate of $10^{11}$ K/s. The inset figure shows the pre-peak (around 1.3 $Å^{-1}$) of the total structure factor. It can be seen that the MD with DNN potential well reproduces the position and height of the pre-peak.